\documentclass[conference,a4paper]{IEEEtran}
\IEEEoverridecommandlockouts

\usepackage{cite}
\usepackage{amsmath,amssymb,amsfonts}
\usepackage{graphicx}
\usepackage{textcomp}
\usepackage{xcolor}
\usepackage{booktabs}
\usepackage{array}
\usepackage{tabularx}
\usepackage{tikz}
\usepackage{multirow}
\usepackage{subcaption}
\renewcommand{\baselinestretch}{0.977}

\def\BibTeX{{\rm B\kern-.05em{\sc i\kern-.025em b}\kern-.08em
    T\kern-.1667em\lower.7ex\hbox{E}\kern-.125emX}}

\begin{document}

\title{Receiver-Centric TDOA Localization Framework for DAB Signals under Synchronization Impairments}

\author{
\IEEEauthorblockN{
Mustafa Furkan Beker, Abdullah Tan, Khaled Walid Elgammal, and Mehmet Kemal Ozdemir
}
\IEEEauthorblockA{
School of Engineering and Natural Sciences, Istanbul Medipol University, Istanbul, Turkey\\
\{mustafa.beker, abdullah.tan\}@std.medipol.edu.tr,
\{khaled.elgammal, mkozdemir\}@medipol.edu.tr
}
}

\IEEEpubid{\makebox[\columnwidth]{\textbf{979-8-3195-1046-4/26/\$31.00~\textcopyright2026 IEEE}\hfill}
\hspace{\columnsep}\makebox[\columnwidth]{}}
\maketitle

\begin{abstract}
In Global Navigation Satellite System (GNSS)-denied environments, terrestrial signals of opportunity (SoOP) offer an alternative for positioning, but synchronization impairments such as clock offsets, drift, and multipath limit performance. This paper proposes a receiver-centric multi-channel time-difference-of-arrival (TDOA) localization framework based on Digital Audio Broadcasting (DAB) signals. The method exploits the DAB null symbol for coarse timing and the phase reference symbol (PRS) for fine synchronization, followed by sub-sample time-of-arrival (TOA) estimation. A double-difference formulation removes inter-receiver clock offsets, while a peak-to-sidelobe ratio (PSR)-based weighting improves robustness. A bias correction step mitigates errors due to multipath. Finally, a coordinated-turn extended Kalman filter (CT-EKF) further refines position estimates. Results show improved accuracy over conventional TDOA with Gauss–Newton estimation, especially in challenging conditions.
\end{abstract}

\begin{IEEEkeywords}
Signals of Opportunity (SoOP), Digital Audio Broadcasting (DAB), Time Difference of Arrival (TDOA), GNSS-Denied Localization, Receiver-Centric Localization
\end{IEEEkeywords}

\section{Introduction}

Global Navigation Satellite System (GNSS)-based positioning systems provide accurate localization under open-sky conditions; however, their performance degrades significantly in urban, indoor, and obstructed environments, where signal blockage, jamming, and spoofing become critical issues \cite{gps_spoofing}. As a result, terrestrial signals of opportunity (SoOP) have emerged as a promising alternative for localization in GNSS-denied environments by leveraging existing communication and broadcast infrastructure \cite{kassas_sop, morales_sop}. Recent studies on terrestrial cellular SoOP further show that existing communication signals can complement GNSS-based positioning through differential and multi-epoch localization formulations \cite{khalife_cd_cellular, zhang_psdm}.

Passive localization using SoOP signals commonly relies on time-based measurements such as time-of-arrival (TOA) and time-difference-of-arrival (TDOA) \cite{torrieri_passive, martin_tdoa}. Broadcast-based systems such as Digital Video Broadcasting--Terrestrial (DVB-T) have demonstrated the feasibility of passive TDOA localization in simulation and experimental studies \cite{dvb_t_exp}. Despite these advances, the use of Digital Audio Broadcasting (DAB) signals for positioning remains relatively limited. Nevertheless, DAB represents an attractive SoOP candidate due to its wide-area coverage, continuous transmission, and deterministic frame structure, which make it a useful terrestrial source in challenging GNSS-denied and mixed-signal localization scenarios. Although DAB provides well-defined synchronization structures such as the null symbol and phase reference symbol (PRS) \cite{etsi_dab}, existing studies have primarily focused on synchronization rather than complete localization frameworks \cite{maul_dab_soo_2023}.

A major challenge in SoOP-based localization is the presence of synchronization impairments, including inter-receiver clock offsets, clock drift, multipath propagation, and non-line-of-sight (NLOS) effects, which introduce significant biases in TOA measurements. Conventional TDOA approaches often rely on simplified assumptions, limiting their applicability in realistic environments \cite{martin_tdoa, clock_sync, multipath_bias}.

To address these challenges, this paper proposes a receiver-centric multi-channel TDOA localization framework based on DAB signals under synchronization impairments. The proposed approach exploits the null symbol for coarse timing and the PRS for fine synchronization, followed by sub-sample TOA estimation. A double-difference formulation eliminates inter-receiver clock offsets, while peak-to-sidelobe ratio (PSR)-based reliability weighting and bias correction improve robustness against distorted correlation peaks, multipath, and NLOS effects. Finally, a coordinated-turn extended Kalman filter (CT-EKF), following the classical Kalman filtering framework \cite{kalman}, is applied for smoothing the trajectory.

\begin{figure}[t]
\centering
\includegraphics[width=\columnwidth]{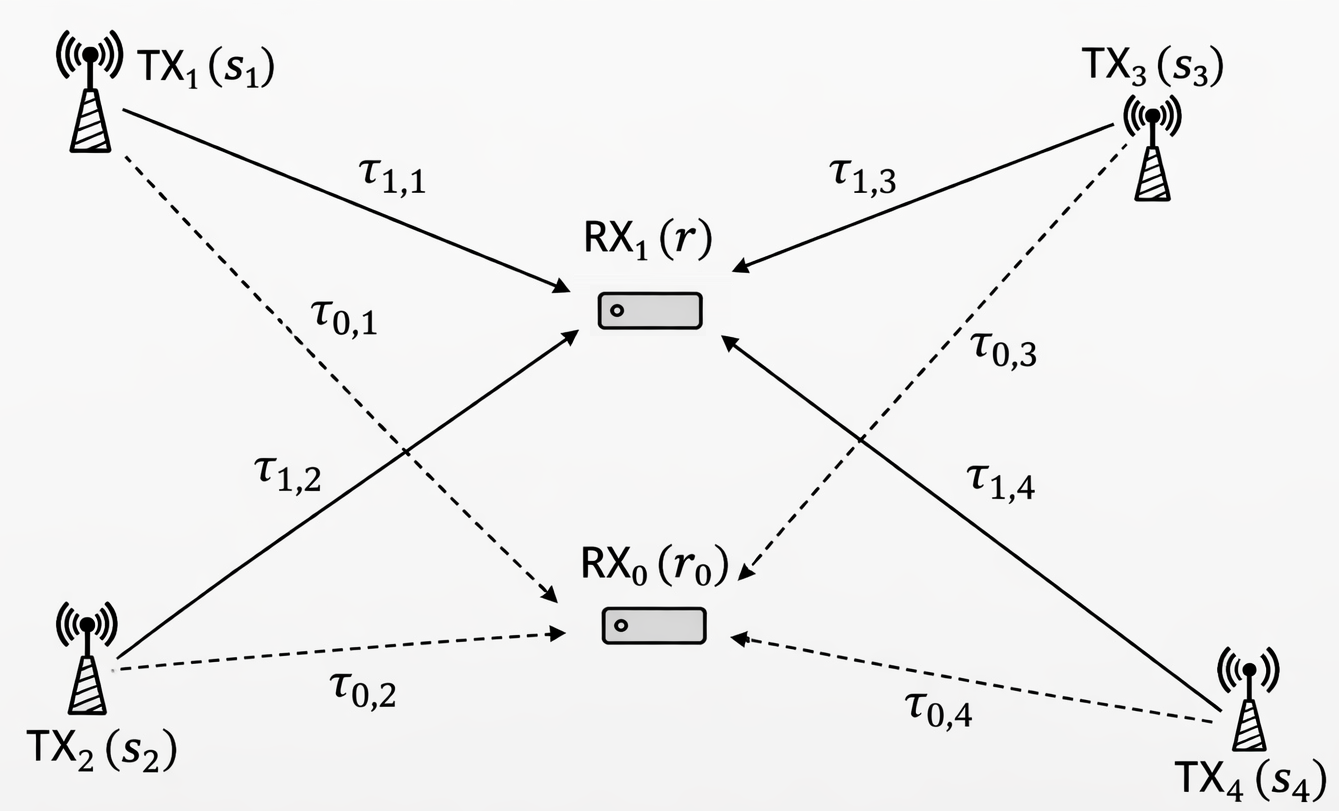}
\caption{System architecture for the proposed method.}
\label{fig:system_model}\vspace{-10pt}
\end{figure}

The main contributions of this work can be summarized as follows: (i) a receiver-centric TDOA localization framework leveraging multi-channel DAB signals under realistic synchronization impairments, (ii) a double-difference formulation combined with PSR-based reliability weighting for robust TDOA-based localization, (iii) a bias correction strategy to compensate systematic timing errors, and (iv) a temporal filtering approach based on CT-EKF to improve tracking stability.

\section{System Model}

The proposed receiver-centric TDOA framework, illustrated in Fig.~\ref{fig:system_model}, exploits DAB signals of opportunity received by two spatially separated receivers: a reference receiver with known position $\mathbf{r}_0$ and a target receiver with unknown position $\mathbf{r}$. The system consists of $M=4$ DAB transmitters with known positions $\mathbf{s}_k=[x_k,y_k]^T$, and aims to estimate the unknown target position $\mathbf{r}$ using receiver-centric TDOA measurements.

The term receiver-centric means that, for each transmitter, the measurement is formed across the two receivers rather than across transmitters. This is suitable for SoOP-based localization, where transmitter synchronization is generally unavailable, as it suppresses common receiver-side timing terms before the double-difference step. The unknown node is modeled as a target receiver in a 2-D formulation.

DAB signals employ an orthogonal frequency-division multiplexing (OFDM) waveform with a deterministic frame structure \cite{etsi_dab}. In the proposed framework, TOA estimation is obtained via null-symbol-based coarse synchronization followed by PRS-based correlation for fine timing, with sub-sample interpolation for improved resolution. The resulting TOA estimates are used to form receiver-centric differential and double-difference measurements, while the PSR of the correlation output serves as a reliability indicator for each channel. The PSR is defined as the ratio between the main correlation peak and the maximum sidelobe level.

The propagation delay between transmitter $k$ and receiver $i \in \{0,1\}$ is given by
\begin{equation}
\tau_{i,k} = \frac{\|\mathbf{r}_i-\mathbf{s}_k\|}{c}.
\end{equation}

In practical conditions, the estimated TOA is affected by multipath, receiver clock offset and drift, carrier frequency offset (CFO), channel-dependent receiver-chain timing bias, and random snapshot jitter. These effects are modeled as
\begin{equation}
t_{i,k}(t) = \tau_{i,k} + b_i(t) + \delta_{i,k}^{\mathrm{rx}} + \nu_{i,k} + \epsilon_{i,k},
\end{equation}
where $b_i(t)=b_{i,0}+\dot{b}_i t$ represents the receiver clock offset with drift, $\delta_{i,k}^{\mathrm{rx}}$ denotes channel-dependent receiver-chain timing bias, $\nu_{i,k}$ models random snapshot jitter, and $\epsilon_{i,k}$ captures residual estimation errors due to noise, multipath distortion, and imperfect synchronization.

To compensate systematic channel-dependent timing errors, a bias correction stage is applied prior to localization. The estimated relative TOA is modeled as
\begin{equation}
\hat{t}_{i,k}^{\mathrm{rel}} =
t_{i,k}^{\mathrm{rel}} + \beta_{i,k} + \epsilon_{i,k},
\end{equation}
where $\beta_{i,k}$ denotes a deterministic channel-dependent bias term caused by receiver hardware imperfections and multipath propagation. The bias is estimated for each receiver-channel pair using prior calibration snapshots and subtracted to obtain
\begin{equation} 
\tilde{t}_{i,k}^{\mathrm{rel}} =
\hat{t}_{i,k}^{\mathrm{rel}} - \hat{\beta}_{i,k}.
\end{equation}
This mainly compensates quasi-static timing offsets, while residual effects are handled through PSR-based weighting.

Receiver-centric differential measurements are formed as
\begin{equation}
\Delta t_k =
\tilde{t}_{1,k}^{\mathrm{rel}} - \tilde{t}_{0,k}^{\mathrm{rel}},
\end{equation}
and a double-difference formulation is applied by selecting transmitter $1$ as the anchor:
\begin{equation}
\delta t_k = \Delta t_k - \Delta t_1, \qquad k=2,\dots,M.
\end{equation}

Using the geometric propagation model, the nonlinear measurement equation becomes
\begin{equation}
\delta t_k =
\frac{1}{c}\left(
\|\mathbf{r}-\mathbf{s}_k\|-\|\mathbf{r}-\mathbf{s}_1\|
-\|\mathbf{r}_0-\mathbf{s}_k\|+\|\mathbf{r}_0-\mathbf{s}_1\|
\right)
+ \eta_k,
\end{equation}
where $\eta_k \sim \mathcal{N}(0, \sigma_k^2)$ represents residual measurement noise. The variance $\sigma_k^2$ is inversely related to the reliability of the correlation peak, quantified via the PSR metric.

\begin{figure}[!t]
\centering
\includegraphics[width=\columnwidth]{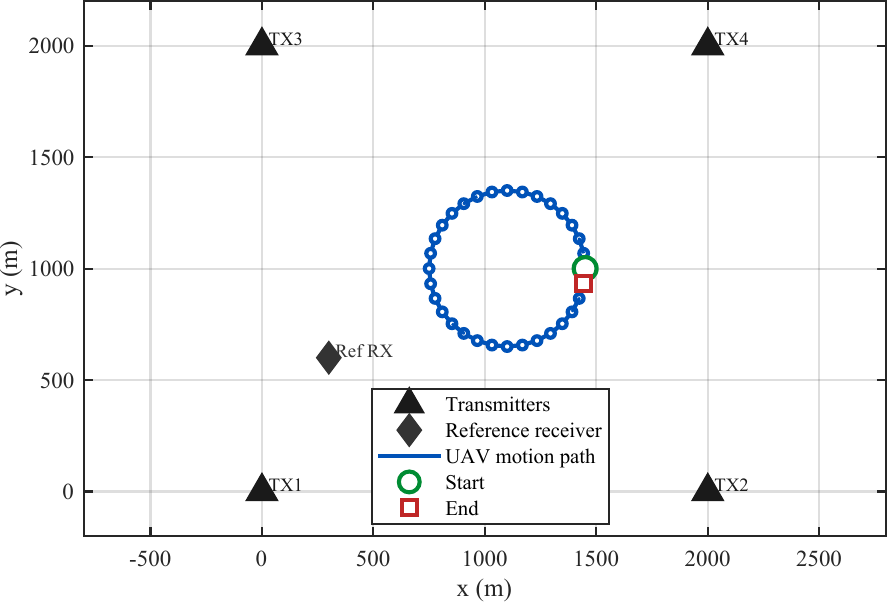}
\caption{Simulation geometry in the considered scenario.}
\label{fig:geometry}\vspace{-10pt}
\end{figure}

The target position is estimated at each snapshot by solving a weighted nonlinear least-squares problem
\begin{equation}
\hat{\mathbf{r}} =
\arg\min_{\mathbf{r}}
\sum_{k=2}^{M}
w_k
\left(
\|\mathbf{r}-\mathbf{s}_k\|-\|\mathbf{r}-\mathbf{s}_1\| - z_k
\right)^2,
\end{equation}
where $z_k = c\,\delta t_k$. The weights $w_k$ are derived from PSR values, assigning lower influence to unreliable channels. The resulting optimization problem is solved iteratively using a weighted Gauss--Newton (GN) algorithm.

To improve temporal consistency, the sequence of GN-based position estimates is filtered using a CT-EKF. The state vector is defined as
\begin{equation}
\mathbf{x}_n = [x_n \;\; y_n \;\; v_n \;\; \theta_n \;\; \omega_n]^T,
\end{equation}
and evolves according to a coordinated-turn motion model
\begin{equation}
\mathbf{x}_{n+1} = f(\mathbf{x}_n) + \mathbf{q}_n.
\end{equation}
For a turn rate $\omega_n \neq 0$, the nonlinear state transition function $f(\mathbf{x}_n)$ is explicitly given by
\begin{equation}
f(\mathbf{x}_n) =
\begin{bmatrix}
x_n + \frac{v_n}{\omega_n} \left( \sin(\theta_n + \omega_n \Delta t) - \sin(\theta_n) \right) \\
y_n + \frac{v_n}{\omega_n} \left( -\cos(\theta_n + \omega_n \Delta t) + \cos(\theta_n) \right) \\
v_n \\
\theta_n + \omega_n \Delta t \\
\omega_n
\end{bmatrix},
\end{equation}
where $\Delta t$ is the sampling interval. When $\omega_n \approx 0$, the model reduces to a constant-velocity model. The GN-based Cartesian position estimate is treated as a noisy observation of the true state:
\begin{equation}
\mathbf{z}_n =
\begin{bmatrix}
\hat{x}^{\mathrm{GN}}_n \\
\hat{y}^{\mathrm{GN}}_n
\end{bmatrix}
=
\mathbf{H}\mathbf{x}_n + \mathbf{v}_n,
\end{equation}
with
\begin{equation}
\mathbf{H}=
\begin{bmatrix}
1&0&0&0&0\\
0&1&0&0&0
\end{bmatrix}.
\end{equation}

In the proposed framework, the CT-EKF is applied in a post-localization manner. The filter predicts the state as
\begin{equation}
\hat{\mathbf{x}}_{n|n-1}=f(\hat{\mathbf{x}}_{n-1|n-1}), \qquad
\mathbf{P}_{n|n-1}=\mathbf{F}_n \mathbf{P}_{n-1|n-1}\mathbf{F}_n^T+\mathbf{Q},
\end{equation}
where $\mathbf{F}_n=\partial f/\partial \mathbf{x}$ is the local Jacobian, and then updates the prediction using the measurement model in (12)–(13) through the standard Kalman measurement update equations, including the Kalman gain computation and covariance correction, thereby reducing snapshot-level noise and improving trajectory smoothness. In this way, the CT-EKF smooths the sequence of snapshot-level GN solutions while preserving the original TDOA-based localization stage.

\section{Simulation Results}

\begin{table}[t]
\caption{Simulation Setup}
\label{tab:sim_setup}
\centering
\setlength{\tabcolsep}{5pt}
\renewcommand{\arraystretch}{1.05}
\begin{tabularx}{\columnwidth}{@{}lX@{}}
\toprule
\textbf{Parameter} & \textbf{Value} \\
\midrule

Sampling rate & 2.048 MS/s (baseband), 40.96 MS/s (wideband) \\
Transmitters & 4 (195.936–206.352 MHz) \\
TX geometry & $(0,0)$, $(2000,0)$, $(0,2000)$, $(2000,2000)$ m \\
Reference receiver & $(300,600)$ m \\

Trajectory & Circular (center $(1100,1000)$ m, radius 350 m) \\
Samples & 1000 positions, $\Delta t = 0.20$ s \\

Burst structure & 3 frames, 20 ms lead / 10 ms tail \\
Synchronization & Null + PRS-based timing \\
TOA estimation & Energy detection + correlation (sub-sample refined) \\

Baseline method & TDOA + GN \\
Proposed method & Receiver-centric bias-corrected TDOA + CT-EKF \\
Measurement model & Double-difference TDOA \\
Weighting & PSR-based \\

Channel cases & Good / Moderate / Harsh \\
SNR (dB) & 28 / 20 / 10 \\
Multipath & 2 / 3 / 5 taps \\
CFO & 0.05–0.75 ppm \\
Clock effects & Offset + drift (case-dependent) \\
Receiver effects & Bias + jitter (case-dependent) \\

Metrics & Mean, P95, max error \\

\bottomrule
\end{tabularx}
\end{table}

The proposed receiver-centric TDOA framework is evaluated under realistic synchronization impairments and channel conditions. The simulation setup is summarized in Table~\ref{tab:sim_setup} and illustrated in Fig.~\ref{fig:geometry}. The scenario includes four DAB transmitters, a known reference receiver, and a moving target receiver following a circular trajectory in a $2~\text{km} \times 2~\text{km}$ area. Timing measurements are obtained using null-symbol-based coarse synchronization and PRS-based fine timing.

Three channel conditions, namely \textit{Good}, \textit{Moderate}, and \textit{Harsh}, are considered using fixed case-dependent SNR, CFO, multipath, clock, and receiver-chain bias settings. Additive white Gaussian noise (AWGN) and per-snapshot timing jitter are modeled as zero-mean Gaussian random variables, and performance is evaluated using mean, P95, and maximum positioning errors.

\begin{table}[t]
\centering
\caption{Localization performance comparison under different channel conditions}
\label{tab:localization_performance}
\setlength{\tabcolsep}{4.5pt}
\renewcommand{\arraystretch}{1.1}
\begin{tabular}{llccc}
\toprule
Case & Method & Mean (m) & P95 (m) & Max (m) \\
\midrule
\multirow{2}{*}{Good}
& Conventional TDOA   & 53.237 & 91.148 & 105.660 \\
& Proposed Method     & 29.354 & 49.322 & 65.559 \\
\midrule
\multirow{2}{*}{Moderate}
& Conventional TDOA   & 63.137 & 117.190 & 128.740 \\
& Proposed Method     & 48.558 & 82.351 & 108.720 \\
\midrule
\multirow{2}{*}{Harsh}
& Conventional TDOA   & 91.097 & 131.500 & 134.980 \\
& Proposed Method     & 74.747 & 108.961 & 124.535 \\
\bottomrule
\end{tabular}\vspace{-5pt}
\end{table}

Across all channel conditions, Table~\ref{tab:localization_performance} shows that the proposed method consistently improves localization performance relative to the conventional TDOA + GN baseline. In the \textit{Good} scenario, the improvement is the most pronounced, with approximately 45\% reduction in mean error, 46\% reduction in P95 error, and 38\% reduction in maximum error. In the \textit{Moderate} scenario, the proposed framework maintains a clear advantage, achieving about 23\% lower mean error, 30\% lower P95 error, and 16\% lower maximum error. Even in the \textit{Harsh} scenario, where synchronization impairments and multipath effects are strongest, the method still provides about 18\% reduction in mean error, 17\% reduction in P95 error, and 8\% reduction in maximum error. Overall, the results indicate that the proposed framework not only improves average localization accuracy but also reduces large-error events, with the largest relative gains observed under more favorable channel conditions and a consistent robustness benefit preserved as the environment becomes more challenging.

\begin{figure}[!t]
\centering
\includegraphics[width=\columnwidth]{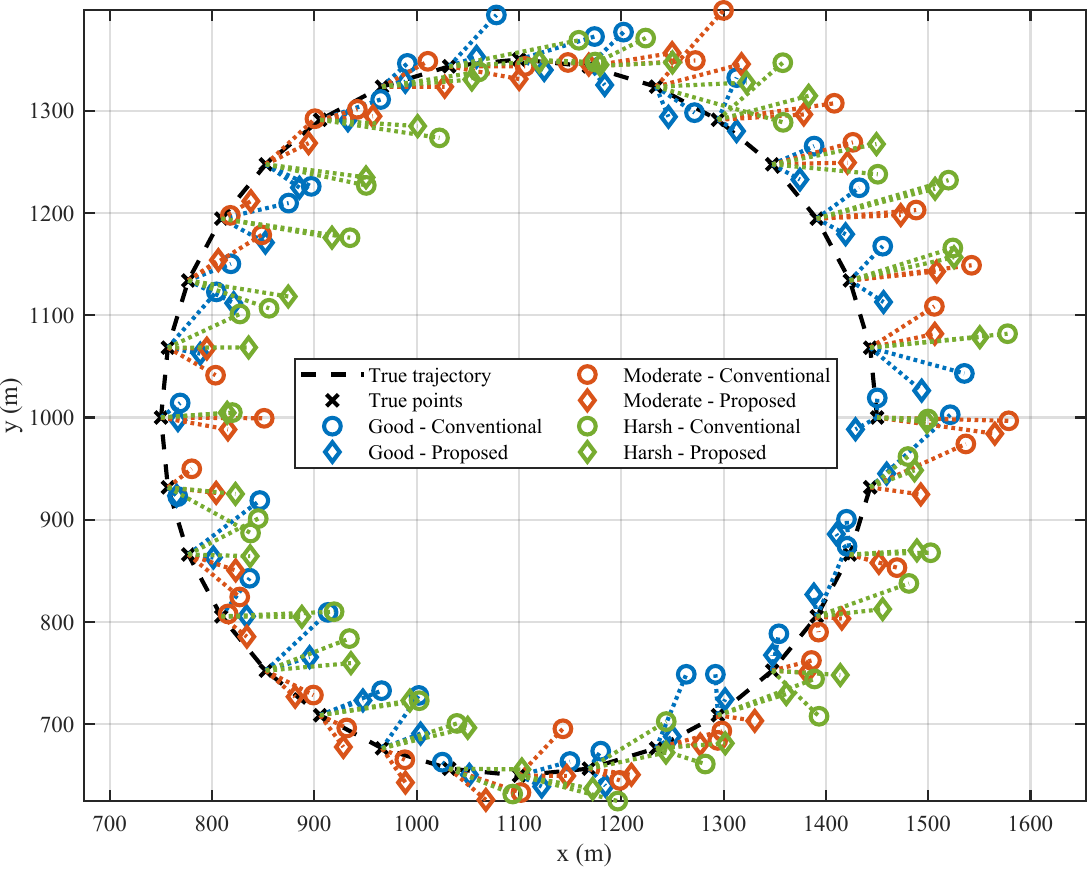}
\caption{True and estimated target receiver positions for conventional and proposed methods.}
\label{fig:traj_compare}\vspace{-12pt}
\end{figure}

\begin{figure*}[!t]
\centering
\includegraphics[width=\textwidth]{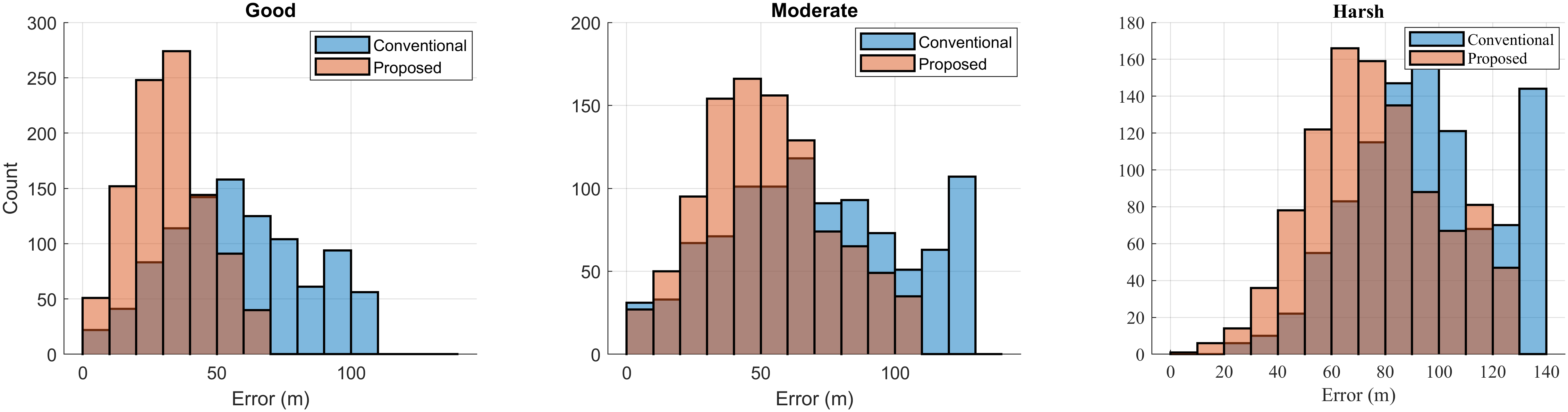}
\caption{Histogram of localization errors for conventional and proposed methods under different channel conditions.}
\label{fig:histogram}\vspace{-12pt}
\end{figure*}

Figure~\ref{fig:traj_compare} illustrates the estimated trajectories obtained using both methods under different channel conditions. As observed, the conventional TDOA + GN solution exhibits a systematic spatial deviation (primarily eastward) rather than being zero-mean random noise. This directional bias can be primarily attributed to the combined effects of network geometry and multipath propagation. Multipath effects inherently cause positive delays, leading to overestimated TOA values. Due to the asymmetric placement of the reference receiver (located at $x=300, y=600$ m) relative to the target trajectory, these positive range errors project unevenly onto the Cartesian domain, pulling the Gauss--Newton solution consistently in one direction governed by the Geometric Dilution of Precision (GDOP). In contrast, the proposed method closely follows the true trajectory, as the bias correction stage effectively isolates and removes this dominant geometric shift, while the temporal filtering smooths residual variations.

Figure~\ref{fig:histogram} shows the histogram of localization errors under different channel conditions. The proposed method shifts the error distribution toward lower values, resulting in a higher concentration of estimates within smaller error ranges. This effect is particularly evident in the moderate and harsh scenarios, where the spread of large errors is significantly reduced.

The observed performance gains can be attributed to three main factors. First, the bias correction stage compensates for systematic timing errors introduced by receiver hardware imperfections, clock offsets, and multipath propagation. Second, the PSR-based weighting mechanism reduces the impact of unreliable channels by assigning lower weights to distorted correlation peaks. Finally, the CT-EKF provides temporal smoothing, reducing fluctuations caused by snapshot-level noise and improving trajectory consistency.

\section{Conclusion}

This paper presented a receiver-centric TDOA localization framework based on multi-channel DAB signals of opportunity for operation in GNSS-denied environments. The proposed approach exploits the DAB frame structure (Null and PRS) for coarse and fine timing, while sub-sample peak refinement improves delay estimation under multipath and low-SNR conditions. A receiver-centric double-difference measurement formulation is employed to suppress inter-receiver clock offsets without requiring absolute synchronization, and the resulting nonlinear localization problem is solved using a GN estimator with temporal consistency enhanced through a CT-EKF. Simulation results demonstrate that the proposed framework consistently outperforms the conventional TDOA approach across good, moderate, and harsh channel conditions, achieving significant reductions in mean, P95, and maximum localization errors and effectively mitigating large-error events. Overall, the results show that multi-channel DAB signals are a promising SoOP source for GNSS-denied localization and that the proposed framework provides a practical and robust solution. DAB is relevant both as a standalone source in limited-availability scenarios and as a candidate signal in mixed-signal opportunistic localization frameworks. Future work will extend the current 2-D formulation to 3-D localization, evaluate asymmetric transmitter deployments and different receiver trajectories, and validate the framework using software-defined radio (SDR)-based DAB measurements.

\section*{Acknowledgment}
This work was supported by TUBITAK under the 1001 project, Grant No. 125E069.

\bibliographystyle{IEEEtran}
\bibliography{references}

\end{document}